\documentclass[11pt, oneside]{article}
\usepackage[margin=3cm]{geometry}
\usepackage{amsmath,amssymb}
\usepackage{graphicx}
\usepackage{float}
\usepackage{psfrag,epsf}
\usepackage{pgf,tikz}
\usepackage{epstopdf}
\usepackage[T1]{fontenc}
\usepackage[toc,page,header]{appendix}
\usepackage{minitoc}
\usepackage{subfiles}
\usepackage{chapterbib}
\RequirePackage{amsthm,amsmath}
\usepackage{natbib}
\usepackage{url}
\usepackage{booktabs}
\usepackage{amsfonts}
\usepackage{nicefrac}
\usepackage{microtype}
\usepackage{multirow}
\usepackage{threeparttable}
\usepackage{siunitx}
\usepackage{caption}
\usepackage{subcaption}
\usepackage{ragged2e}
\usepackage{hyperref}
\usepackage{authblk}
\hypersetup{colorlinks,linkcolor={blue},citecolor={blue},urlcolor={blue}}

\definecolor{myblue}{rgb}{0,0,0.75}

\usepackage{lineno}
\makeatletter
\def\makeLineNumberLeft{%
  \linenumberfont\llap{\hb@xt@\linenumberwidth{\LineNumber\hss}\hskip\linenumbersep}%
  \hskip\columnwidth%
  \rlap{\hskip\linenumbersep\hb@xt@\linenumberwidth{\hss\LineNumber}}\hss}
\leftlinenumbers
\makeatother

\title{VA-Calibration: Correcting for Algorithmic Misclassification in Estimating Cause Distributions}

\author[1]{Sandipan Pramanik\thanks{Corresponding Author. Email: \href{mailto:spraman4@jhu.edu}{spraman4@jhu.edu}. ORCiD: \href{https://orcid.org/0000-0002-7196-155X}{0000-0002-7196-155X}}}
\author[2]{Emily B. Wilson\thanks{Email: \href{mailto:ewilso28@jhu.edu}{ewilso28@jhu.edu}}}
\author[2]{Henry D. Kalter\thanks{Email: \href{mailto:hkalter1@jhu.edu}{hkalter1@jhu.edu}}}
\author[2]{Agbessi Amouzou\thanks{Email: \href{mailto:aamouzo1@jhu.edu}{aamouzo1@jhu.edu}}}
\author[2]{Robert E. Black\thanks{Email: \href{mailto:rblack1@jhu.edu}{rblack1@jhu.edu}}}
\author[3]{Li Liu\thanks{Email: \href{mailto:lliu26@jhu.edu}{lliu26@jhu.edu}}}
\author[2]{Jamie Perin\thanks{Email: \href{mailto:jperin@jhu.edu}{jperin@jhu.edu}}}
\author[1]{Abhirup Datta\thanks{Email: \href{mailto:abhidatta@jhu.edu}{abhidatta@jhu.edu}}}

\affil[1]{Department of Biostatistics, Johns Hopkins University}
\affil[2]{Department of International Health, Johns Hopkins University}
\affil[3]{Department of International Health and Department of Population, Family and Reproductive Health, Johns Hopkins University}

\date{\ }

\begin{document}

\maketitle

\begin{abstract}
Accurate estimation of cause-specific mortality fractions (CSMFs), the percentage of deaths attributable to each cause in a population, is essential for global health monitoring. Challenge arises because computer-coded verbal autopsy (CCVA) algorithms, commonly used to estimate CSMFs, frequently misclassify the cause of death (COD). This misclassification is further complicated by structured patterns and substantial variation across countries. To address this, we introduce the R package \href{https://cran.r-project.org/package=vacalibration}{\texttt{vacalibration}} \citep{Pramanik2026package}. It implements a modular Bayesian framework to correct for the misclassification, thereby yielding more accurate CSMF estimates from verbal autopsy (VA) questionnaire data.

The package utilizes uncertainty-quantified CCVA misclassification matrix estimates derived from data collected in the CHAMPS project and available on the \href{https://github.com/sandy-pramanik/CCVA-Misclassification-Matrices}{CCVA-Misclassification-Matrices} GitHub repository. Currently, these matrices cover three CCVA algorithms (EAVA, InSilicoVA, and InterVA) and two age groups (neonates aged 0-27 days, and children aged 1-59 months) across countries (specific estimates for Bangladesh, Ethiopia, Kenya, Mali, Mozambique, Sierra Leone, and South Africa, and a combined estimate for all other countries), enabling global calibration. The \href{https://cran.r-project.org/package=vacalibration}{\texttt{vacalibration}} package also supports ensemble calibration when multiple algorithms are available.

Implemented using the `RStan', the package offers rapid computation, uncertainty quantification, and seamless compatibility with \href{https://cran.r-project.org/package=openVA}{\texttt{openVA}} R package, a leading COD analysis software ecosystem \citep{openva2023, openva2024}. We demonstrate the package's flexibility with two real-world applications in \href{https://comsamozambique.org/}{COMSA-Mozambique} and \href{https://childmortality.org/about}{CA CODE}. The package and its foundational methodology apply more broadly and can calibrate any discrete classifier or its ensemble.
\end{abstract}


\section{Introduction}\label{intro}

Understanding the disease burden in a population is fundamental to effective public health practice. A key measure in this regard is the cause-specific mortality fractions (CSMF). They represent the proportion of deaths in a population attributable to different causes, which are derived by aggregating individual-level causes of death (CODs). However, complete and reliable COD from complete diagnostic autopsies (CDA) is scarce in many low- and middle-income countries {[}LMICs; \citet{Nichols2018}{]}. As a result, alternative approaches such as minimally invasive tissue sampling (MITS) and verbal autopsy (VA) have gained prominence. The COD determination using MITS has demonstrated high agreement with that using CDA \citep{bassat2017, menendez2017}. However, due to its resource-intensive nature, it is typically limited to select hospitals and is not scalable to entire populations.

VA, in contrast, is a less-invasive method in which caregivers are systematically interviewed to gather the decedent's medical history and symptoms \citep{mikkelsen2015, wang2017, adair2018}. Computer-coded VA (CCVA) algorithms then process the resulting information to assign an individual-level COD. The automation and scalability of CCVA algorithms have made VA the leading method for determining CODs in community settings, where it is routinely used to obtain population-level CSMF estimates \citep{setel2005, soleman2006, garenne2014, Savigny2017}. This approach has been applied in the Countrywide Mortality Surveillance for Action (COMSA) programs in Mozambique (\href{https://comsamozambique.org/}{COMSA-Mozambique}) and Sierra Leone, where nationwide VA studies produced mortality estimates across all age groups to monitor progress toward the health-related Sustainable Development Goals \citep{Marsh2022, perin2022, Fiksel2023, Gilbert2023, Macicame2023}.

A wide range of specialized CCVA algorithms has been developed to predict COD from VA records. Examples include EAVA {[}Expert Algorithm; \citet{kalter2016}; \citet{Wilson2025}{]}, InSilicoVA \citep{McCormick2016}, InterVA \citep{byass2012}, Tariff \citep{james2011}, the King and Lu method \citep{King2008}, and more recently, domain adaptation-based approaches \citep{wu2024tree, li2024bayesian}. Many of these tools have been consolidated into the \href{https://cran.r-project.org/package=openVA}{\texttt{openVA}} R package \citep{openva2023, openva2024}. In addition, general-purpose machine learning methods, such as random forests \citep{breiman2001random}, naive Bayes classifiers \citep{Minsky1961}, and support vector machines \citep{cortes1995support}, have also been applied to VA data \citep{Flaxman2011, Miasnikof2015, Koopman2015}. The compatibility of the WHO-standardized VA questionnaire with these algorithms has enabled the widespread adoption of VA for COD determination and estimation of national and sub-national CSMFs in many LMICs.

Despite their wide applicability, CCVA algorithms are prone to algorithmic misclassification, wherein discrepancies between algorithm-assigned and true CODs can introduce bias into CSMF estimates. Moreover, the misclassifications are often systematic and vary across countries, algorithms, and age groups, leading to incorrect population-level mortality patterns \citep{Pramanik2025aoas, Pramanik2026bmjgh}. Correcting such systematic errors is therefore critical for producing valid and comparable COD estimates across countries and time periods.

To address this, we present \href{https://cran.r-project.org/package=vacalibration}{\texttt{vacalibration}} \citep{Pramanik2026package}, an R package that implements statistically principled frameworks and analysis workflows for correcting algorithmic misclassification in estimating population-level CSMFs from VA data. \citet{Pramanik2026bmjgh} recently analyzed limited paired COD data from the Child Health and Mortality Prevention Surveillance (\href{https://champshealth.org/}{CHAMPS}) project. For each death in the study, the CODs were determined by two methods: a VA, and a gold-standard diagnosis using MITS \citep{blau2019}. They have published country-specific CCVA misclassification matrix estimates on the \href{https://github.com/sandy-pramanik/CCVA-Misclassification-Matrices}{CCVA-Misclassification-Matrices} GitHub repository, which cover three widely used CCVA algorithms (EAVA, InSilicoVA, and InterVA) and two age groups (neonates aged 0-27 days; children aged 1-59 months) across countries (specific estimates for Bangladesh, Ethiopia, Kenya, Mali, Mozambique, Sierra Leone, and South Africa, and a combined estimate for all other countries). Leveraging this, the \href{https://cran.r-project.org/package=vacalibration}{\texttt{vacalibration}} package accepts CCVA-derived COD as input, and employs the misclassification matrix estimates as informative priors in a modular Bayesian model to produce uncertainty-quantified, calibrated CSMF estimates.

The methodology underlying \href{https://cran.r-project.org/package=vacalibration}{\texttt{vacalibration}} employs a Bayesian calibration model that accounts for uncertainty in estimating CCVA misclassification and population-level CSMFs. Moreover, its modular design enables calibration without requiring direct access to labeled COD data, such as from \href{https://champshealth.org/}{CHAMPS}. This protects the privacy of the labeled data while leveraging knowledge about CCVA misclassification derived from them. \href{https://cran.r-project.org/package=vacalibration}{\texttt{vacalibration}} supports multiple algorithms, directly accepts outputs from CCVA algorithms (currently EAVA, InSilicoVA, and InterVA) as inputs, and seamlessly integrates with \href{https://cran.r-project.org/web/packages/openVA/vignettes/vacalibration-vignette.html}{openVA}, which is a leading VA-based COD analysis workflow. This allows users to calibrate VA-only data reproducibly and consistently across studies. The package is publicly available on \href{https://cran.r-project.org/package=vacalibration}{CRAN},
and is actively maintained on \href{https://github.com/sandy-pramanik/vacalibration}{GitHub}.

The remainder of this article is organized as follows. Section \ref{intro-comsamoz} introduces the VA data collected in \href{https://comsamozambique.org/}{COMSA-Mozambique}, which serves as the motivating application. Section \ref{intro-champs} describes the labeled COD data from the \href{https://champshealth.org/}{CHAMPS} project, which is used to estimate CCVA misclassification matrices. Section \ref{method} presents the statistical framework that underpins the package, while Section \ref{package-overview} provides an overview of the key functions in the package. Section \ref{implement} demonstrates the use of the package through two real-world applications: single-country calibration using \href{https://comsamozambique.org/}{COMSA-Mozambique} data and multi-country calibration in the Child and Adolescent Causes of Death Estimation (\href{https://childmortality.org/about}{CA CODE}) project. Finally, Section \ref{discussion} concludes with a summary of findings and future directions.

\section{Example dataset: COMSA-Mozambique verbal autopsy records}\label{intro-comsamoz}

The use of VA to create nationally representative COD databases is exemplified by the Countrywide Mortality Surveillance for Action (COMSA) initiatives in Mozambique (\href{https://comsamozambique.org/}{COMSA-Mozambique}) and Sierra Leone \citep{Marsh2022, Macicame2023}. These initiatives routinely utilize the databases to generate age-specific national and sub-national CSMF estimates \citep{Fiksel2023, Gilbert2023, Macicame2023}.

As an illustrative example, this article focuses on \href{https://comsamozambique.org/}{COMSA-Mozambique}, which was established to generate representative COD estimates across all age groups \citep{Macicame2023}. Provincially representative pregnancy and mortality data were collected through routine community surveillance in clusters, each consisting of hundreds of households. Reported deaths were followed by caregiver interviews using an integrated verbal and social autopsy questionnaire adapted from the \href{https://www.who.int/publications/m/item/verbal-autopsy-standards-the-2016-who-verbal-autopsy-instrument}{2016 WHO VA instrument}. The VA survey for each individual is then passed through one or multiple CCVA algorithms to obtain individual-level CODs, and is grouped into six broad causes for neonates and nine for children. Neonatal causes included congenital malformation, pneumonia, sepsis/meningitis/infections, intrapartum-related events (IPRE), prematurity, and `other'. Child causes encompassed malaria, pneumonia, diarrhea, severe malnutrition, HIV/AIDS, injury, neonatal causes (IPRE, congenital malformation, prematurity), other infections, and `other'. These are referred to as the (unlabeled) VA COD data, as the true CODs for them are unknown.

Let \(\mathcal{U} := {\cal U}_s\) denote the VA-only data \(\{{\boldsymbol{V}}_r\}_{r \in {\cal U}}\) of \(N\) deaths from a target country \(s\), where \(\boldsymbol{V}_r = \left( V_{r1},\dots,V_{rC} \right)^\top\) is the estimated VA-COD for an individual \(r\), and \(V_{rj}=1\) if its VA-COD is \(j\). In VA-only studies such as \href{https://comsamozambique.org/}{COMSA-Mozambique}, VA-CODs are routinely aggregated to obtain the raw or uncalibrated estimate of CSMF (\(q_j\)) for cause \(j\) as
\begin{equation}\label{eq: uncalib}
\hat{q}_j = \frac{1}{N} \sum_{r \in \mathcal{U}} V_{rj} = \frac{\text{Number of VA-predicted deaths from cause } j}{\text{Total number of VA records}}.
\end{equation}
However, compared to medical certification, full autopsy, or MITS, CCVA algorithms frequently misclassify true COD \citep{datta2020, fiksel2022}. This makes the uncalibrated CSMF deviate from its true values. This is described by the calibration equation:
\begin{equation}\label{eq: calibeq}
q_j = \sum_{i=1}^C \phi_{ij} p_i \quad \Leftrightarrow \quad \boldsymbol{q} = \Phi^\top \, \boldsymbol{p},
\end{equation}
where \(C\) is the total number of causes, \(p_i\) is the true CSMF for cause \(i\), \(\Phi = (\phi_{ij})\) is the \(C \times C\) misclassification matrix, and \(\phi_{ij} = \mathbb{P} \left( V=j \mid M=i \right)\) denotes the probability of the algorithm \((V)\) assigning cause \(j\) when the gold standard diagnosis \((M)\) assigns cause \(i\) (\(\phi_{ii}\) on the diagonal are sensitivities, and \(\phi_{ij}\) on the off-diagonals represent false negatives).

\section{Characterizing algorithmic misclassification using CHAMPS data}\label{intro-champs}

As the number of deaths (\(N\)) in a VA-COD database increases, it yields uncalibrated CSMF estimates that are highly precise. However, this precision is misleading as it ignores the misclassification of CCVA algorithms, which in turn leads to biased and overconfident results. To address this, VA-Calibration employs a modular Bayesian framework \citep{datta2020, fiksel2022, Pramanik2025aoas}. First, it accounts for the systematic pattern and cross-country heterogeneity of CCVA misclassification, and estimates misclassification matrices using limited paired COD data that includes both VA and a gold standard diagnosis (such as MITS). Then it solves an inverse problem to correct for the misclassification, yielding a calibrated CSMF estimate. While this increases uncertainty, it improves out-of-sample predictive performance.

For estimating CCVA misclassification, we leverage labeled COD data collected in the Child Health and Mortality Prevention Surveillance (\href{https://champshealth.org/}{CHAMPS}) project, which is an ongoing initiative surveilling child mortality in Africa and South Asia \citep{blau2019, Salzberg2019}. \href{https://champshealth.org/}{CHAMPS} employs a rapid mortality notification system, ensuring that all deaths among stillbirths and children under age five are reported within 24 hours to local teams. Following notification, informed consent is obtained from parents or guardians for eligible cases. The COD is then investigated through MITS and a comprehensive suite of laboratory analyses. These analyses include microbiological and histopathological testing, and diagnostic testing for HIV, tuberculosis, malaria, and other infectious agents. Additionally, caregiver interviews conducted via VA are requested to gather detailed symptom histories and contextual clinical information. A multidisciplinary Determination of Cause of Death (DeCoDe) panel subsequently reviews all collected data to ascertain the COD, which is then communicated to the family. In instances of multiple causes, the DeCoDe panel establishes the causal chain, identifying the primary (or underlying) and secondary (or immediate and intermediate) causes. For this study, we utilize only the primary cause, referred to as the ``CHAMPS cause''. \citet{Pramanik2026bmjgh} recently analyzed CCVA misclassification using CHAMPS data. The deaths included in the analysis occurred between December 2016 and June 2023 in Bangladesh, Ethiopia, Kenya, Mali, Mozambique, Sierra Leone, and South Africa. The CODs were categorized into the broad causes previously mentioned. Given the high concordance of CHAMPS COD with CDA COD \citep{Byass2016, bassat2017, blau2019}, we utilize the CHAMPS causes as the gold standard for evaluating the misclassification of CCVA algorithms. See \citet{Pramanik2026bmjgh} for more details.

\section{Method}\label{method}

In this section, we present methodologies underlying the \href{https://cran.r-project.org/package=vacalibration}{vacalibration} package that are employed to derive the CCVA misclassification matrix estimates and to obtain the calibrated CSMF estimates in this package. In recent work, \citet{Pramanik2025aoas} introduced a comprehensive framework for modeling country-specific misclassification matrices of CCVA algorithms, designed specifically for settings with limited labeled data. The framework underlies a parsimonious model for estimating a homogeneous misclassification matrix, which is then extended to model country-specific matrices using a hierarchical Bayesian setup. The framework accommodates cross-country variation and employs data-driven shrinkage to favor simpler models when sample sizes are insufficient or evidence is weak.

Additionally, they proposed a modular Bayesian framework for calibrating VA-only COD data (e.g., \href{https://comsamozambique.org/}{COMSA-Mozambique}), which does not require direct access to the labeled COD data (e.g., \href{https://champshealth.org/}{CHAMPS}). The framework first analyzes the labeled data using the misclassification matrix model, and uncertainty-quantified estimates of the misclassification matrices are obtained and stored. Subsequently, these estimates are integrated as informative priors within a Bayesian hierarchical model for the downstream task of estimating CSMFs. Below, we summarize key components of the two frameworks.

\subsection{Modeling country-specific misclassification matrices}\label{missmat-model}

\subsubsection{Base model: Parsimonious homogeneous modeling of misclassification matrices}\label{base-model}

The parsimonious misclassification matrix model captures two fundamental behaviors of algorithms: correct identification of the CHAMPS cause by design, and a systematic bias toward specific causes when an algorithm fails to identify the CHAMPS cause.

\paragraph{Intrinsic accuracy (diagonal effect).} CCVA algorithms are fundamentally designed to assign COD as accurately as possible. Let \(a_i\) represent the \textit{intrinsic accuracy} of the algorithm for cause \(i\), that is, the probability that the algorithm correctly predicts the CHAMPS cause \(i\). Intuitively, the intrinsic accuracy reflects the presence of a distinctive symptom pattern in the VA record that arises only when the CHAMPS cause is \(i\).

\paragraph{Systematic preference or pull (column effect).} When an algorithm fails to identify the CHAMPS cause by design, the base model assumes a simplified mechanism whereby it assigns one of the \(C\) causes according to prespecified probabilities, independent of the CHAMPS cause. This captures the algorithm's \textit{systematic preference} or \textit{pull}, toward specific causes. The pull toward cause \(j\) is denoted by \(\rho_j\), and represents the probability of assigning cause \(j\) regardless of the CHAMPS cause when it fails to make a correct assignment.\\

Together, intrinsic accuracy and pull capture the core mechanism underlying the CCVA misclassification. Although neither process is directly observable, the misclassification rates that we observe arise from their joint effects. In the base model, a correct classification for CHAMPS cause \(i\) can occur in two ways: either directly with probability \(a_i\), or through its pull toward cause \(i\) with probability \((1-a_i)\rho_i\) when it fails to match by design. Conversely, misclassification into a different cause \(j \neq i\) occurs only when the designed match fails and the algorithm is pulled toward \(j\), with probability \((1-a_i)\rho_j\). Thus, under the base model, sensitivities and false negatives are given by \(\phi_{ii} = a_i + (1-a_i)\rho_i\) and \(\phi_{ij} = (1-a_i)\rho_j\).

The base model parsimoniously represents homogeneous misclassification using only \(2C-1\) parameters (\(C\) intrinsic accuracies and \(C-1\) pulls) compared to \(C^2 - C\) parameters in the case of no structure. This significantly reduces dimension and is especially important when modeling country-specific effects with limited data. Additionally, by modeling pull, it quantifies systematic biases in algorithm behavior, highlighting causes that are consistently over- or underpredicted. This yields more reliable misclassification estimates and informs improvements in CCVA algorithm design.

\subsubsection{General homogeneous misclassification matrices}\label{general-homogeneous-model}

Although the base model's assumptions ensure parsimony, they are often overly restrictive. Accordingly, a more general homogenous model is presented, which allows for an unstructured, data-driven estimation when sufficient data is available, yet it shrinks toward the base model in the absence of data.

For an algorithm, let \(\{(\boldsymbol{V}_r, \boldsymbol{M}_r)\}_{r \in \mathcal{L}_s}\) denote the paired VA- and CHAMPS-COD in country \(s\). Define the observed misclassification count matrix in country \(s\) as
\begin{equation*}
\text{T}_s = (t_{sij}) = \sum_{r \in \mathcal{L}_s} \, \mathbb{I}\left( {\boldsymbol{V}}_r = j \mid {\boldsymbol{M}}_r = i \right),
\end{equation*}
and the pooled misclassification count matrix be \(\text{T}= \sum_{s=1}^S \text{T}_s\). For cause \(i\), denote the number of cases with CHAMPS cause \(i\) in country \(s\) as \(n_{si} = \sum_{r \in {\cal L}_s} \mathbb{I}(M_r = i),\) and the total number of such cases across all \href{https://champshealth.org/}{CHAMPS} sites as \(n_i = \sum_{s=1}^S n_{si}\).

Suppose \(\Phi = (\phi_{ij})\) denote a general misclassification matrix, and \(\Phi^{(b)} = (\phi^{(b)}_{ij})\) denote the matrix under the base model, with \(\phi^{(b)}_{ii} = a_i + (1-a_i)\rho_i\) and \(\phi^{(b)}_{ij} = (1-a_i)\rho_j\). For modeling purpose, false negative rates in \(\Phi\) reparameterized as \(\phi_{ij} = (1-\phi_{ii}) q_{ij}\) for \(i \neq j\), where \(\phi_{ii}\) is the sensitivity and \(q_{ij} = \mathbb{P}(V=j \mid M=i, V \neq i)\) is the relative false negative rate. By construction, \(q_{ij} > 0\), \(\sum_{j \neq i} q_{ij} = 1\), and it simplifies to \(q^{(b)}_{ij} = \rho_j / (1-\rho_i)\) under the base model.

With \(\text{T}_{i*} = (t_{i1}, \dots, t_{iC})^\top\) and \(\Phi_{i*} = (\phi_{i1}, \dots, \phi_{iC})^\top\) respectively denoting the \(i^{th}\) rows of \(\text{T}\) and \(\Phi\), the observed misclassification counts are hierarchically modeled as
\begin{equation}\label{eq: pooled}
\begin{split}
\text{T}_{i*} & \overset{ind}{\sim} \text{Multinomial}(n_i, \Phi_{i*}), \text{  with  } \phi_{ij}=(1-\phi_{ii}) q_{ij} \quad \forall i \neq j,\\
\phi_{ii} &\overset{ind}{\sim} \text{Beta}\left( 0.5 + 2\omega_P \phi^{(b)}_{ii}, 0.5 + 2\omega_P (1 - \phi^{(b)}_{ii}) \right) \text{  with  } \omega_P>0,\\
{\boldsymbol{q}}_{i*} &\overset{ind}{\sim} \text{Dirichlet}\left( 0.5 + (C-1)\omega_P {\boldsymbol{q}}^{(b)}_{i*} \right) \text{  with  } \omega_P>0,\\
a_i &\overset{iid}{\sim} \text{Beta}\left(b,d\right) \text{  with  } b,d>0,\\
{\boldsymbol{\rho}} &\sim \text{Dirichlet}\left( {\boldsymbol{e}} \right) \text{  with  } {\boldsymbol{e}} = \left(e_1,\dots,e_C\right)^\top \text{ and } e_j>0.
\end{split}
\end{equation}
Note that \eqref{eq: pooled} simplifies to the base model as \(\omega_P \to \infty\). This enhances estimation accuracy under limited data. As \(\omega_P \to 0\), \eqref{eq: pooled} approaches \(\text{Beta}(0.5,0.5)\), \(\text{Dirichlet}(0.5,\dots,0.5)\), the Jeffreys non-informative priors commonly used in modeling proportions. The parameter \(\omega_P\) controls the pull strength towards the base model. By tuning \(\omega_P\), the framework flexibly spans from a structured misclassification with systematic preference to a general misclassification with no structure.

\subsubsection{County-specific misclassification matrices}\label{country-specific-model}

To account for cross-country heterogeneity, country-specific misclassification matrices are modeled hierarchically, centered around the homogeneous matrix:
\begin{equation}\label{eq: het}
\begin{split}
\text{T}_{si*} &\overset{ind}{\sim} \text{Multinomial}(n_{si},\Phi_{si*}) \text{  with  } \phi_{sij}=(1-\phi_{sii}) q_{sij} \quad \forall i,\\
\phi_{sii} &\overset{ind}{\sim} \text{Beta}\left( 0.5 + 2\omega_S \phi_{ii}, 0.5 + 2\omega_S (1- \phi_{ii})\right) \quad \forall i \text{  with  } \omega_S>0, \\
{\boldsymbol{q}}_{si *} &\overset{ind}{\sim} \text{Dirichlet}\left( 0.5 + (C-1)\omega_R {\boldsymbol{q}}_{i *} \right) \quad \forall i \text{  with  } \omega_R>0.
\end{split}
\end{equation}
This is combined with \eqref{eq: pooled} to complete the hierarchical specification. \eqref{eq: het} generalizes the homogeneous model \eqref{eq: pooled}, reducing to the latter when \(\omega_S, \omega_R \to \infty\). Conversely, as \(\omega_S, \omega_R \to 0\), the country-specific sensitivities and relative false negatives follow the Jeffreys non-informative priors corresponding to independent modeling of misclassification in each country without information sharing. The parameters \(\omega_S\) and \(\omega_R\) control degrees of heterogeneity in sensitivity and relative false negatives. \(\omega_P\), \(\omega_S\), and \(\omega_R\) are effect sizes of the framework, and independent shrinkage priors are assumed on them. This yields a data-adaptive trade-off between parsimony and flexibility.

\subsubsection{Misclassification for unobserved CHAMPS causes and countries}

For countries observed in the labeled data, misclassification estimates are given by the posterior distribution from \eqref{eq: pooled}-\eqref{eq: het}. By virtue of Bayesian hierarchical modeling, the framework also predicts misclassification rates for countries or CHAMPS causes that are not observed in the labeled data. The predictive distribution of misclassification rates in a new country \(l\) for CHAMPS cause \(i\) is given by
\begin{equation}\label{eq: pred}
\begin{split}
\phi_{lii} &\sim \text{Beta} \left( 0.5 + 2\omega_S \phi_{ii}, 0.5 + 2\omega_S (1 - \phi_{ii} ) \right), \quad \forall i, \\
\phi_{lij} &= (1 - \phi_{lii}) q_{lij}, \quad \forall i \neq j, \\
{\boldsymbol{q}}_{li*} &\sim \text{Dirichlet} \left( 0.5 + (C-1)\omega_R {\boldsymbol{q}}_{i*} \right).
\end{split}
\end{equation}
Posterior samples from \eqref{eq: pooled}-\eqref{eq: het} are utilized to obtain predictive samples from \eqref{eq: pred}. The predictive distribution is centered at the homogeneous misclassification, and its uncertainty is governed by posteriors of \(\omega_S\) and \(\omega_R\). While point predictions from homogeneous and heterogeneous models are often similar, the heterogeneous model produces wider uncertainty that reflects potential cross-country variation.

\subsection{Modular VA-Calibration with country-specific misclassification matrix}\label{modular-vacalib}

A primary objective in improving CCVA misclassification estimation is to enhance the accuracy of CSMF estimation when calibrating unlabeled VA-only COD data, such as that from \href{https://comsamozambique.org/}{COMSA-Mozambique}. Let \({\cal U}:= {\cal U}_s\) denote the VA-only data from target country \(s\), and let \({\boldsymbol{p}}:= {\boldsymbol{p}}_{{\cal U}}\) denote the corresponding CSMF to be estimated. The approach first fits the framework in Section \ref{missmat-model} to the labeled data (\({\cal L}\)), retaining posterior and predictive samples of country-specific CCVA misclassification matrices. These samples are then used to construct informative priors for calibrating VA-only data, approximated via CHAMPS cause-specific Dirichlet distributions. This leads to the following hierarchical model for VA-Calibration in country \(s\):
\begin{equation}\label{eq: modular calib}
\begin{split}
V_r &\overset{iid}{\sim} \text{Multinomial}(1, {\boldsymbol{\Phi}}_s^\top {\boldsymbol{p}}), \quad \forall r \in \mathcal{U}, \\
{\boldsymbol{\Phi}}_{si*} &\overset{ind}{\sim} \text{Dirichlet}({\boldsymbol{e}}_{si}), \quad \forall i, \quad {\boldsymbol{e}}_{si} = (e_{si1},\dots,e_{siC})^\top, \; e_{sij} > 0, \\
{\boldsymbol{p}}&\sim \text{Dirichlet}(1 + C \eta \hat{{\boldsymbol{q}}}), \quad \eta > 0.
\end{split}
\end{equation}
Here, \({\boldsymbol{e}}_{si}\) are Dirichlet scale parameters chosen to approximate the marginal posterior \([{\boldsymbol{\Phi}}_{si*} {\,|\,}\mathcal{L}]\) in \eqref{eq: pooled}-\eqref{eq: het} in the upstream analysis. The prior on \({\boldsymbol{p}}\) is structured to shrink towards the uncalibrated CSMF estimate \(\hat{{\boldsymbol{q}}}\), as defined in \eqref{eq: uncalib}, when data are limited. The parameter \(\eta\) controls the degree of shrinkage and plays a role analogous to the effect sizes \(\omega_P, \omega_S\), and \(\omega_R\). In this work, we fix \(\eta=4\) following \citet{Pramanik2025aoas}.

The effectiveness of VA-Calibration relies on the transportability assumption; that is, the CCVA misclassification observed in the labeled data \({\cal L}\) is representative of those in the unlabeled data \({\cal U}\) \citep{fiksel2022}. Although this assumption cannot be directly verified due to the absence of true labels in \({\cal U}\), it is supported by the notion that the misclassification primarily depends on the conditional distribution of symptoms given a COD. Because symptom profiles for a given cause tend to be relatively universal, misclassification patterns are expected to be broadly consistent across countries, with heterogeneity arising mainly from differences in disease dynamics or VA implementation practices.

\section{Key objects and functions in the package}\label{package-overview}

\subsection{\texorpdfstring{\texttt{CCVA\_missmat}: CCVA misclassification matrices based on CHAMPS}{CCVA\_missmat: CCVA misclassification matrices based on CHAMPS}}\label{ccva_missmat-ccva-misclassification-matrices-based-on-champs}

Using limited labeled data from \href{https://champshealth.org/}{CHAMPS} and the framework in Section \ref{missmat-model}, \citet{Pramanik2026bmjgh} has produced uncertainty-quantified misclassification matrix estimates for three CCVA algorithms (EAVA, InSilicoVA, and InterVA) and two age groups (neonates and children) across countries (specific estimates for Bangladesh, Ethiopia, Kenya, Mali, Mozambique, Sierra Leone, and South Africa, and a combined estimate for all other countries). They are publicly available on the \href{https://github.com/sandy-pramanik/CCVA-Misclassification-Matrices}{CCVA-Misclassification-Matrices} GitHub repository and also included in \href{https://cran.r-project.org/package=vacalibration}{\texttt{vacalibration}} package. Posterior means and Dirichlet approximations are available in both. Due to the file size limit, posterior samples are only available on the GitHub repository, while the package includes the posterior summary. For example, the posterior mean, Dirichlet approximation, and posterior summary of EAVA's misclassification matrix for neonates in Mozambique can be accessed in the package as

\begin{verbatim}
# install `vacalibration` from CRAN or GitHub
library(vacalibration)                             # load package
CCVA_missmat$neonate$eava$postmean$Mozambique      # posterior mean
CCVA_missmat$neonate$eava$asDirich$Mozambique      # Dirichlet approximation
CCVA_missmat$neonate$eava$postsumm$Mozambique      # posterior summary
CCVA_missmat$neonate$eava$postsamples$Mozambique   # posterior samples
\end{verbatim}

\noindent Estimates for children, other algorithms (InSilicoVA and InterVA), and countries can be accessed similarly (see the documentation of \texttt{CCVA\_missmat} in the R package for details).

\subsection{\texorpdfstring{\texttt{cause\_map}: CCVA output to broad cause}{cause\_map: CCVA output to broad cause}}\label{causemap}

This function accepts outputs of CCVA algorithms (e.g., EAVA, InSilicoVA, InterVA, physician-coded verbal autopsy (PCVA)) as inputs, and maps their individual-level COD assignments to broad causes defined in Section \ref{intro-comsamoz}. In particular, \texttt{cause\_map()} accepts outputs from \texttt{getTopCause} or \texttt{getIndivProb} in \href{https://cran.r-project.org/package=openVA}{\texttt{openVA}} package, and from \texttt{codEAVA} in \href{https://cran.r-project.org/package=EAVA}{\texttt{EAVA}} package. By reducing the effective number of causes from the International Classification of Diseases (ICD) to broad-cause-level, \texttt{cause\_map} standardizes definitions of COD across CCVA algorithms, and improves uncertainty in calibrating CSMF estimates.

As mentioned in Section \ref{intro-comsamoz}, the broad causes for neonates (0-27 days) are `congenital malformation', `pneumonia', `sepsis and meningitis and infections', `intrapartum-related events (IPRE)', `other', and `prematurity'. In children (1-59 months), the broad causes are `malaria', `pneumonia', `diarrhea', `severe malnutrition', `HIV', `injury', `other', `other infections', and a combined `neonatal causes' category which includes `congenital malformation', `IPRE', and `prematurity'.

For illustration, top COD assignments for the de-identified \href{https://comsamozambique.org/}{COMSA-Mozambique} data analyzed in Section \ref{implement-comsamoz} are included in the \href{https://cran.r-project.org/package=vacalibration}{\texttt{vacalibration}} package as a list \texttt{comsamoz\_CCVAoutput}. For example, COD predicted by EAVA are available as \texttt{comsamoz\_CCVAoutput\$neonate\$eava}, which can then be mapped to broad causes as

\begin{verbatim}
######## load library
library(vacalibration)

######## mapping specific causes to broad causes
cause_map_EAVAoutput <- cause_map(df=comsamoz_CCVAoutput$neonate$eava,
                                  age_group="neonate")
\end{verbatim}

\noindent The resulting output is a matrix with individuals as rows and broad causes as columns. For each individual, the broad cause indicates a 1, while all other causes have a 0. This can be implemented similarly for the algorithms in \href{https://cran.r-project.org/package=openVA}{\texttt{openVA}} and ``\texttt{children}''.

\subsection{\texorpdfstring{\texttt{vacalibration}: Calibrating CSMF estimates from CCVA outputs}{vacalibration: Calibrating CSMF estimates from CCVA outputs}}\label{vacalibration-calibrating-csmf-estimates-from-ccva-outputs}

Using outputs from CCVA algorithms together with uncertainty-quantified, country-specific misclassification matrices, this function implements the modular VA-Calibration described in Section \ref{modular-vacalib} and produces national-level CSMF estimates. It calibrates VA-only data by adjusting for misclassification either stored in \texttt{CCVA\_missmat} or user-specified. Below, we summarize its key features.

\paragraph{Input.}

The data can be input via the argument \texttt{va\_data}. It accepts CCVA outputs from \href{https://cran.r-project.org/package=openVA}{\texttt{openVA}} or \href{https://cran.r-project.org/package=EAVA}{\texttt{EAVA}} packages as

\begin{verbatim}
# A tibble
   ID cause      
<dbl> <chr>
11224 Sepsis     
13674 Intrapartum
 3868 Pneumonia
\end{verbatim}

\noindent or outputs from \texttt{cause\_map()} as

\begin{verbatim}
# A binary matrix
congenital_malformation pneumonia sepsis_meningitis_inf ipre other prematurity
                      0         0                     1    0     0           0
                      0         0                     0    1     0           0
                      0         1                     0    1     0           0
\end{verbatim}

\noindent or cause-specific death counts as

\begin{verbatim}
# A named numeric vector
congenital_malformation pneumonia sepsis_meningitis_inf ipre other prematurity
                     44       168                   267  268    29         164 
\end{verbatim}

\noindent Data from multiple algorithms can be input as a named list with component names indicating algorithms. To enable \texttt{vacalibration()} to retrieve the corresponding misclassification matrix from \texttt{CCVA\_missmat}, users are required to specify the desired age group and country using the \texttt{age\_group} and \texttt{country} arguments, respectively. Users can also provide custom misclassification matrices through \texttt{missmat\_type} and \texttt{missmat} as point estimates, random samples, or row-specific Dirichlet priors (Dirichlet scale parameters). In this case, specifying \texttt{age\_group} and \texttt{country} is not required.

\paragraph{Key features.}

The function \((i)\) calibrates using either the stored (\texttt{CCVA\_missmat}) or user-supplied misclassification matrices; \((ii)\) propagates uncertainty in the misclassification matrix via Dirichlet priors; and \((iii)\) enables ensemble calibration across multiple CCVA algorithms. More generally, the package allows us to calibrate population-level proportions derived from categorical classifiers.

\paragraph{Mapping non-CHAMPS broad causes.}

When custom data in \texttt{va\_data} are provided at the broad-cause level (e.g., outputs from \texttt{cause\_map()} or cause-specific death counts) and the observed causes do not align with CHAMPS broad causes, \texttt{vacalibration()} requires a mapping between observed and CHAMPS broad causes to utilize the CHAMPS-based misclassification estimates. This mapping needs to be specified via the \texttt{studycause\_map} argument (e.g., for neonates, \texttt{c(``studycause1''=``pneumonia'', ``studycause2''=``congenital malformation'', ``studycause3''=``pneumonia'', ...)}). This allows the function to construct misclassification matrix estimates from the CHAMPS-based estimates stored in \texttt{CCVA\_missmat}. Alternatively, users may supply custom misclassification estimates directly via the \texttt{missmat} argument.

\paragraph{Uncalibrated causes.}

Algorithm-specific causes that should not be calibrated can be specified via \texttt{donotcalib}. This is a named list analogous in structure to \texttt{va\_data} and \texttt{missmat}, which identifies causes to exclude for each algorithm. By default, the package does not calibrate for VA cause named as `other'. In addition, when misclassification rates for a VA cause do not vary across CHAMPS causes (i.e., little variation across rows within a column), the calibration equation \eqref{eq: calibeq} becomes ill-conditioned \citep[see footnote on p.~1227 in][]{Pramanik2025aoas}. By setting \texttt{donotcalib\_type = ``learn"} (default), the function detects VA causes (columns) with a range of values less than \texttt{nocalib.threshold} (default 0.1). It then appends them to \texttt{donotcalib} alongside any user-specified exclusions. For VA-Calibration, misclassification matrices are normalized row-wise for causes to be calibrated.

\paragraph{Shrinking misclassification matrix for stable calibration.}

To improve stability in VA-Calibration, \citet{datta2020} proposed shrinking the calibrated CSMF toward their uncalibrated estimate \(\hat{{\boldsymbol{q}}}\). Noting this is equivalent to shrinking the misclassification matrix toward the identity matrix, \texttt{vacalibration()} also includes an alternative that sets \texttt{path\_correction=TRUE} and shrinks the misclassification matrix in a data-driven manner. Specifically, given an initial point estimate \(\hat{\Phi}\) of misclassification matrix (such as the posterior mean in \texttt{CCVA\_missmat}), it defines a path of misclassification matrices \(\hat{\Phi}_\lambda = \lambda \textbf{I} + (1-\lambda)\hat{\Phi}\) for \(\lambda \in [0,1]\) (e.g., grid values), where \(\textbf{I}\) is the \(C \times C\) identity matrix. It then chooses the maximum value of \(\lambda\), \(\hat{\lambda}\), such that \(\hat{\Phi}_\lambda^{-\top} \, \hat{{\boldsymbol{q}}}\) lies strictly inside the \(C\)-dimensional simplex. VA-Calibration then implements \eqref{eq: modular calib} by fixing \(\eta = 0\) and using \(\hat{\Phi}_{\hat{\lambda}}\) as the misclassification matrix estimate. The resulting degree of shrinkage (\(\hat{\lambda}\)) is available with the output as \texttt{lambda\_calibpath}. This can be disabled by setting \texttt{path\_correction=FALSE}, in which case it implements \eqref{eq: modular calib} with specified \(\eta\) via \texttt{pshrink\_strength} (default 4).

\paragraph{Outputs from VA-Calibration.}

\texttt{vacalibration()} returns a comprehensive list of outputs, including posterior samples (\texttt{p\_calib}) and summaries (\texttt{pcalib\_postsumm}) of calibrated CSMFs, uncalibrated CSMFs (\texttt{p\_uncalib}), calibrated (\texttt{va\_deaths\_calib\_algo}) and uncalibrated (\texttt{va\_deaths\_uncalib}) cause-specific death counts. The output also contains misclassification matrices (\texttt{Mmat\_tomodel}) or their row-specific Dirichlet priors used for calibration, together with logical indicators (\texttt{donotcalib\_tomodel}) specifying causes that are not calibrated (e.g., `other'). Further details with example illustrations are provided below.

\section{Implementation examples}\label{implement}

This section demonstrates how the package can be integrated into existing VA-based mortality surveillance workflows to produce national-level calibrated CSMF estimates. Using data from \href{https://comsamozambique.org/}{COMSA-Mozambique}, Sections \ref{odk-to-ccva}-\ref{implement-comsamoz} discuss a complete workflow from raw VA questionnaire responses to CSMF estimation. To illustrate applicability in a different context, Section \ref{implement-cacode} presents a second application in \href{https://childmortality.org/about}{CA CODE}, where only aggregated cause-specific death counts from multiple countries are available, and the causes observed across studies do not align with the CHAMPS broad causes for which the CCVA misclassification matrix estimates are stored in \texttt{CCVA\_missmat}.

\subsection{Estimating individual-level top causes of death from raw ODK VA records}\label{odk-to-ccva}

In Mozambique, VA-based mortality surveillance provides an electronic, near-real-time framework for data collection, review, analysis, and dissemination. Using the Open Data Kit (ODK) platform on smartphones, community surveillance agents record pregnancy outcomes and deaths, and conduct verbal and social autopsies for reported deaths. Data are stored on cloud servers and analyzed using statistical software, such as \texttt{R} and \texttt{Stata} (see \href{https://viva.jhuhost.org/}{Vital Insights for Vital Action}).

\paragraph{Mapping raw ODK VA records for CCVA implementation.}

CCVA algorithms require input VA data in the standard WHO 2016 VA questionnaire format. An example data in this format, \texttt{\"who151\_odk\_export.csv\"}, is provided in the \href{https://cran.r-project.org/package=openVA}{\texttt{openVA}} package and is used here for illustration. In \texttt{R}, the data can be loaded as

\begin{figure}[b]

{\centering \includegraphics[width=0.98\linewidth]{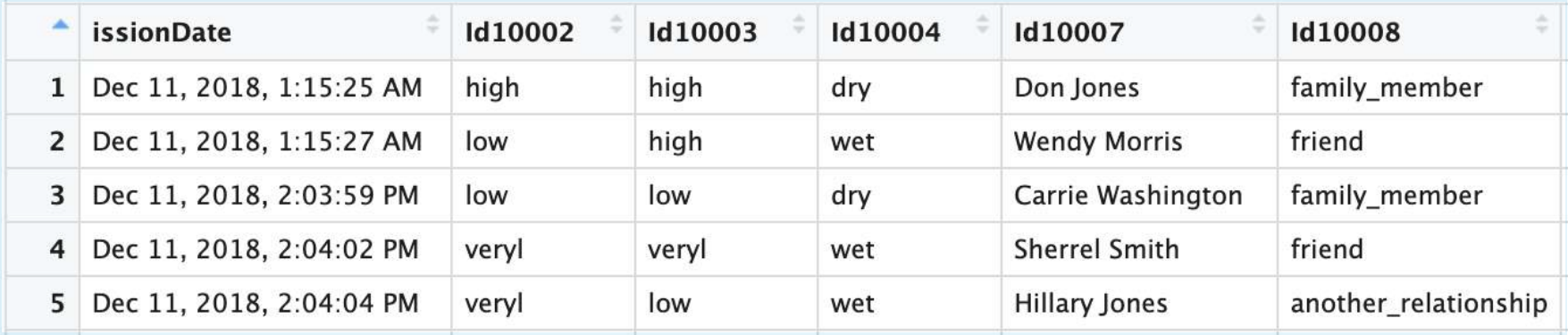} 

}

\caption{Example VA survey dataset (\texttt{records}) in the standard WHO 2016 VA questionnaire format, provided as \texttt{"who151\_odk\_export.csv"} in the \texttt{CrossVA} package.}\label{fig:exampleva-fig}
\end{figure}

\begin{verbatim}
######## load libraries
library(openVA)
library(CrossVA)

######## read in sample data
example_file <- system.file("sample", "who151_odk_export.csv", 
                            package = "CrossVA")
records <- read.csv(example_file, stringsAsFactors = FALSE)

######## reformat some variables for EAVA
names(records) <- sub(".*[.]?(?i)(id\\d{5}.*)", "\\1", names(records), perl=TRUE)
names(records) <- sub(".*age", "age", names(records))
names(records) <- sub(".*display", "display", names(records))
names(records) <- sub(".*is", "is", names(records))
colnames(records)
records$ageindaysnew <- records$ageInDays
\end{verbatim}

\noindent The data are structured as a data frame, with individuals as rows, and questions about symptoms and other attributes as columns. Figure \ref{fig:exampleva-fig} shows a snapshot of the example data.

To implement CCVA algorithms, the raw VA responses are passed through \texttt{odk2EAVA} in the \href{https://cran.r-project.org/package=EAVA}{\texttt{EAVA}} package for EAVA, or \texttt{odk2openVA\_v151} in the \href{https://cran.r-project.org/package=openVA}{\texttt{openVA}} package for algorithms present there. The functions convert each VA response, including character variables and symptom durations, into binary indicators (\texttt{\"yes\"}, \texttt{\"no\"}, or missing \texttt{\".\"}). For the example data \texttt{records}, this transformation for EAVA is performed as

\begin{figure}[t]

{\centering \includegraphics[width=0.98\linewidth]{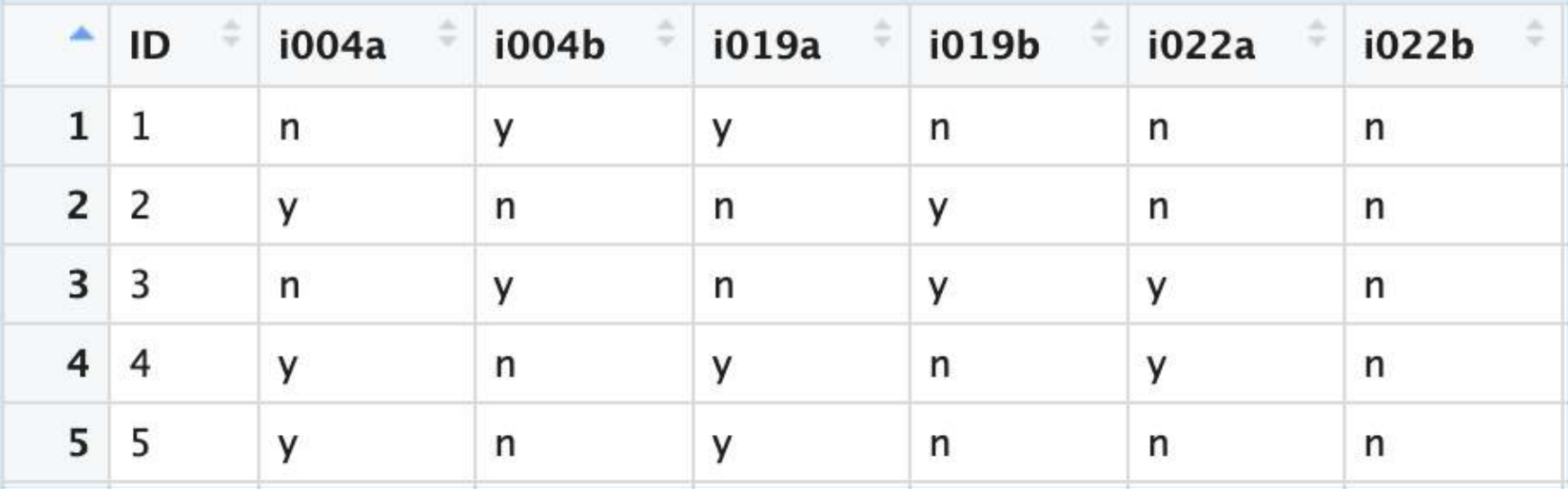} 

}

\caption{These are VA survey data (\texttt{records}) in Figure \ref{fig:exampleva-fig} that are mapped using \texttt{odk2EAVA} for input to the EAVA algorithm.}\label{fig:example-mapped-eava-fig}
\end{figure}
\begin{figure}[t]

{\centering \includegraphics[width=0.98\linewidth]{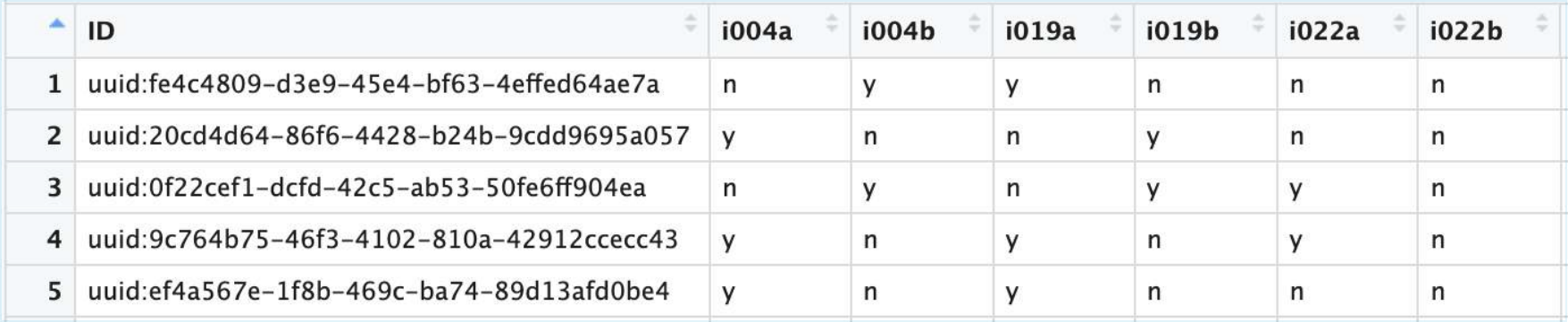} 

}

\caption{These are VA survey data (\texttt{records}) in Figure \ref{fig:exampleva-fig} that are mapped using \texttt{odk2openVA\_v151} for input to algorithms in `openVA` (e.g., InSilicoVA, InterVA).}\label{fig:example-mapped-openva-fig}
\end{figure}

\begin{verbatim}
######## load EAVA package
library(EAVA)
library(stringi)
library(stringr)

######## run odk2EAVA() using example data
mapped_records_EAVA <- odk2EAVA(records, "KEY")
\end{verbatim}

\noindent For InSilicoVA and InterVA, this is done in the \href{https://cran.r-project.org/package=openVA}{\texttt{openVA}} as

\begin{verbatim}
######## run CrossVA to map variables to openVA algorithm input format
mapped_records_openVA <- odk2openVA_v151(records, id_col = "KEY")
\end{verbatim}

\noindent Snapshots of the mapped data for input to \href{https://cran.r-project.org/package=EAVA}{EAVA} (\texttt{mapped\_records\_EAVA}) and \href{https://cran.r-project.org/package=openVA}{\texttt{openVA}} (\texttt{mapped\_records\_openVA}) are presented in Figure \ref{fig:example-mapped-eava-fig} and \ref{fig:example-mapped-openva-fig}. By input to CCVA algorithms, we obtain individual-level COD assignments using \href{https://cran.r-project.org/package=EAVA}{\texttt{EAVA}} and \href{https://cran.r-project.org/package=openVA}{\texttt{openVA}}. Below, we illustrate this workflow in the \href{https://comsamozambique.org/}{COMSA-Mozambique} application.

\paragraph{Applying CCVA algorithms to obtain individual-level top COD.}

After obtaining the mapped records, they can be input into \href{https://cran.r-project.org/package=EAVA}{\texttt{EAVA}} and \href{https://cran.r-project.org/package=openVA}{\texttt{openVA}} to implement the CCVA algorithms and obtain the top COD as follows.

\begin{verbatim}
######## EAVA algorithm
EAVAoutput <- codEAVA(df = mapped_records_EAVA, age_group = "neonate")

######## InSilicoVA algorithm
get_InSilicoVA_CODs <- codeVA(mapped_records_openVA, data.type = "WHO2016", 
                              model = "InSilicoVA", version = "5.0",
                              HIV = "h", Malaria = "h")
InSilicoVAoutput <- getTopCOD(get_InSilicoVA_CODs)

######## InterVA algorithm
get_InterVA5_CODs <- codeVA(mapped_records_openVA, data.type = "WHO2016",
                            model = "InterVA", version = "5",
                            HIV = "h", Malaria = "h", write=FALSE)
InterVA5output <- getTopCOD(get_InterVA5_CODs)
\end{verbatim}

\subsection{Implementing VA workflow in COMSA-Mozambique}\label{implement-comsamoz}

The Countrywide Mortality Surveillance for Action project in Mozambique (\href{https://comsamozambique.org/}{COMSA-Mozambique}) is a nationally representative, continuous mortality surveillance system. It is designed to generate timely estimates of mortality and COD at national and provincial levels for use by government and policy-makers. Mozambique, an East African country with a population of roughly 35 million, is one of the few sub-Saharan African countries to have achieved Millennium Development Goal 4 by reducing under-five mortality by more than two-thirds by 2015 (to an estimated 79 deaths per 1,000 live births). Meeting the Sustainable Development Goal (SDG) target of under-five mortality below 25 per 1,000 live births by 2030 will require strengthened health programs and precise, up-to-date measurement of mortality and COD patterns. To facilitate this, we aim to utilize the \href{https://cran.r-project.org/package=vacalibration}{\texttt{vacalibration}} package to obtain more valid national-level CSMF estimates in Mozambique for broad CODs. More generally, the estimation workflow can serve as a model for generating VA-based CSMF estimates in any country.

The VA workflow to obtain calibrated CSMF estimates from raw VA questionnaire responses is implemented in two steps.

\paragraph{Step 1. Obtaining the top COD from raw ODK VA records.}

Given raw individual-level responses to the VA questionnaire, we follow the steps in Section \ref{odk-to-ccva} and obtain individual-level top COD. A de-identified VA survey data from \href{https://comsamozambique.org/}{COMSA-Mozambique} used in our application is available \href{https://comsamozambique.org/data-access}{here} by registering and accepting the Data Use Agreement (DUA). Unlike the sample data used in Section \ref{odk-to-ccva} for illustration, VA survey records collected in field settings in practice, including \href{https://comsamozambique.org/}{COMSA-Mozambique}, often deviate from the standard WHO 2016 VA questionnaire format. Consequently, some preprocessing and adaptation are required before they can be input into \texttt{odk2EAVA} or \texttt{odk2openVA\_v151}. The specific troubleshooting steps are highly dependent on the data collection instrument and implementation context. Comprehensive descriptions of VA data processing and the application of CCVA algorithms are available in resources such as \citet{byass2020}, \citet{openva2023}, and \citet{Wilson2025}.

Once appropriately harmonized into the required format, the data are passed to \texttt{odk2EAVA} and \texttt{odk2openVA\_v151}, as demonstrated in Section \ref{odk-to-ccva}, to obtain binary-encoded VA records \texttt{mapped\_records\_EAVA} and \texttt{mapped\_records\_openVA}. The mapped records are then input into the \href{https://cran.r-project.org/package=EAVA}{\texttt{EAVA}} and \href{https://cran.r-project.org/package=openVA}{\texttt{openVA}} to implement the CCVA algorithms. The top COD assignments for the de-identified \href{https://comsamozambique.org/}{COMSA-Mozambique} data analyzed here are included with the \href{https://cran.r-project.org/package=vacalibration}{\texttt{vacalibration}} package as the \texttt{list} object \texttt{comsamoz\_CCVAoutput} for three algorithms (EAVA, InSilicoVA, and InterVA) and two age groups (neonates and children). For example, the output from EAVA for neonates can be accessed as \texttt{comsamoz\_CCVAoutput\$neonate\$eava}.

\paragraph{Step 2. VA-Calibration using CCVA outputs.}

The CCVA outputs obtained above (for example, \texttt{comsamoz\_CCVAoutput\$neonate\$eava}) are then input into \texttt{vacalibration()} via \texttt{va\_data}. For example, VA-Calibration for EAVA among neonates is implemented as

\begin{verbatim}
######## load library
library(vacalibration)

######## VA-Calibration with output from EAVA
va_data_eava <- list("eava"=comsamoz_CCVAoutput$neonate$eava)
vacalib_eava <- vacalibration(va_data=va_data_eava, age_group="neonate",
                              country="Mozambique")
\end{verbatim}

\noindent This can be similarly implemented for other algorithms and age groups.

\paragraph{Ensemble VA-Calibration.}

If outputs from multiple CCVA algorithms are available, they can be supplied jointly via \texttt{va\_data} as a named list, with components and their names corresponding to algorithms. With \texttt{ensemble=TRUE} (default), \texttt{vacalibration()} performs an ensemble calibration in addition to algorithm-specific calibrations. For example, for neonates in \href{https://comsamozambique.org/}{COMSA-Mozambique}, this can be implemented as:

\begin{verbatim}
######## Ensemble VA-Calibration with outputs from EAVA, InSilicoVA, and InterVA
va_data_ensemble <- list("eava"=comsamoz_CCVAoutput$neonate$eava,
                         "insilicova"=comsamoz_CCVAoutput$neonate$insilicova,
                         "interva"=comsamoz_CCVAoutput$neonate$interva)
                        
vacalib_ensemble <- vacalibration(va_data=va_data_ensemble,
                                  age_group="neonate", country="Mozambique")
\end{verbatim}

\paragraph{Interpreting outputs from VA-Calibration.}

Algorithm-specific calibration provides a list of outputs (e.g., \texttt{vacalib\_eava}), including the uncalibrated point estimate of CSMFs (\texttt{p\_uncalib}), posterior samples of calibrated CSMF (\texttt{p\_calib}), and their posterior summary (\texttt{pcalib\_postsumm}; mean and 95\% credible interval):

\begin{verbatim}
vacalib_eava$p_uncalib[1,] # uncalibrated CSMF estimates
vacalib_eava$p_calib[1,,] # posterior samples of calibrated CSMF
vacalib_eava$pcalib_postsumm[1,,] # posterior summary
\end{verbatim}

\noindent The output also provides uncalibrated and calibrated death counts available as \texttt{va\_deaths\_uncalib} and \texttt{va\_deaths\_calib\_algo}:

\begin{verbatim}
vacalib_eava$va_deaths_uncalib[1,] # uncalibrated counts
vacalib_eava$va_deaths_calib_algo[1,] # calibrated counts
\end{verbatim}

\noindent The output from ensemble calibration \texttt{vacalib\_ensemble} is also similarly structured. Here is a summary of key outputs.

\begin{verbatim}
######## uncalibrated CSMF estimates
vacalib_ensemble$p_uncalib

######## calibrated CSMF estimates
vacalib_ensemble$p_calib["eava",,]         # EAVA
vacalib_ensemble$p_calib["insilicova",,]   # InSilicoVA
vacalib_ensemble$p_calib["interva",,]      # InterVA
vacalib_ensemble$p_calib["ensemble",,]     # Ensemble

######## posterior summaries
vacalib_ensemble$pcalib_postsumm["eava",,]         # EAVA
vacalib_ensemble$pcalib_postsumm["insilicova",,]   # InSilicoVA
vacalib_ensemble$pcalib_postsumm["interva",,]      # InterVA
vacalib_ensemble$pcalib_postsumm["ensemble",,]     # Ensemble

######## cause-specific death counts
vacalib_ensemble$va_deaths_uncalib          # uncalibrated
vacalib_ensemble$va_deaths_calib_algo       # algorithm-specific calibration
vacalib_ensemble$va_deaths_calib_ensemble   # ensemble calibration
\end{verbatim}

\begin{figure}[!t]

{\centering \includegraphics[width=0.95\linewidth]{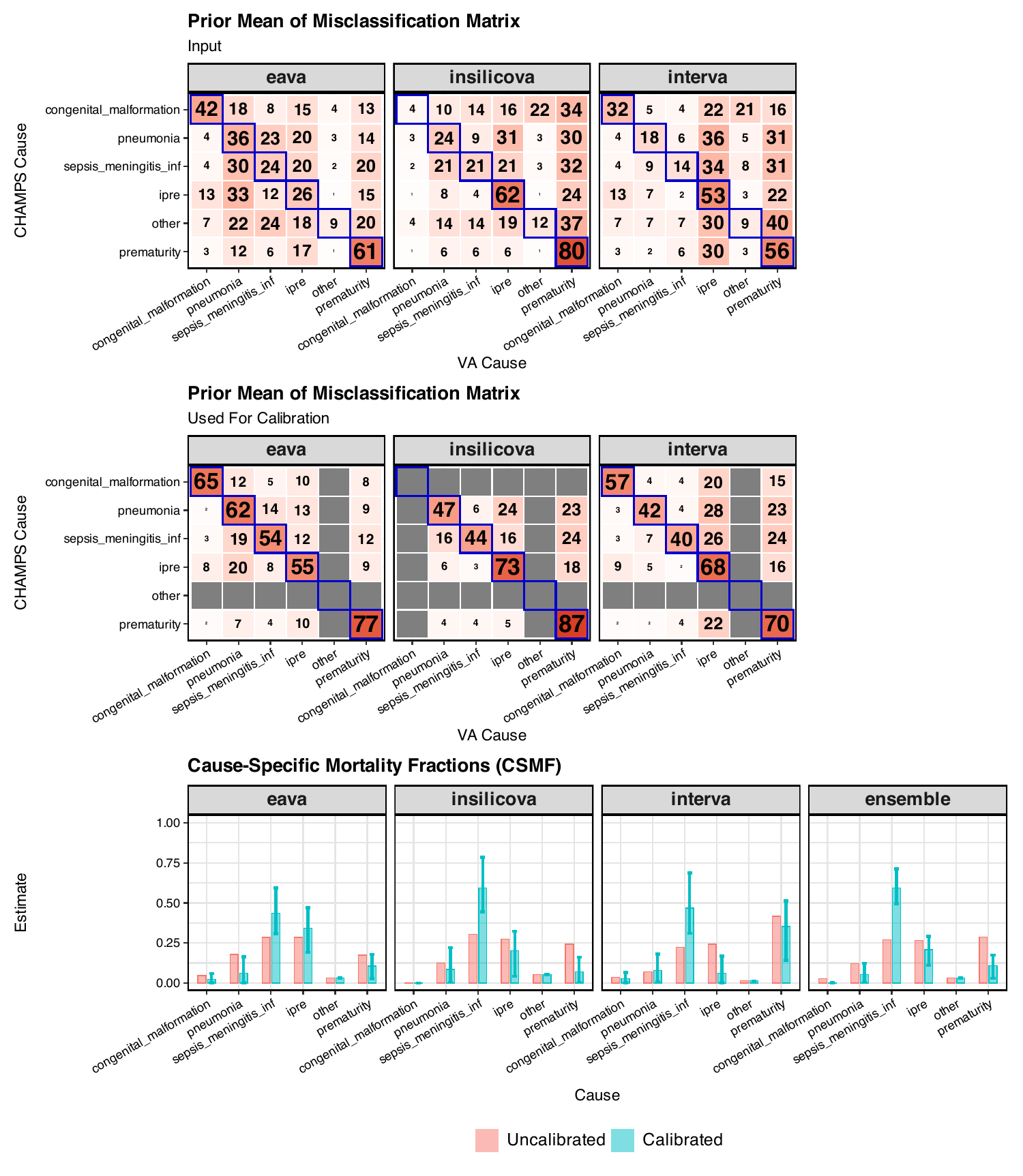} 

}

\caption{The top panel displays the average CHAMPS-based misclassification matrix estimates for each algorithm as stored. The middle panel shows the average misclassification matrix after correction. This is used for calibration in an uncertainty-quantified informative prior. The bottom panel compares the uncalibrated CSMF estimates with the corresponding uncertainty-quantified calibrated CSMF estimates. Grey rows and columns in the misclassification matrix indicate causes that are not calibrated.}\label{fig:comsamoz-ensemble-plot}
\end{figure}

\paragraph{Visualization.}

After VA-Calibration is conducted, the output from \texttt{vacalibration()} can be directly passed to \texttt{plot\_vacalib()} to generate a plot summarizing the main components of VA-Calibration. This plot displays the misclassification matrix used for calibration, and presents both the uncalibrated and calibrated CSMF estimates. The specific misclassification matrix plotted depends on the \texttt{missmat\_type} specified in \texttt{vacalibration()}. If \texttt{missmat\_type=\"fixed\"}, the fixed misclassification matrix used for calibration is plotted. If \texttt{missmat\_type} is set to \texttt{\"prior\"} or \texttt{\"samples\"}, the average misclassification matrix is displayed. For example, when calibrating for EAVA as demonstrated above, the summary plot can be generated as:

\begin{verbatim}
plot_vacalib(vacalib_fit = vacalib_eava,
             toplot = "missmat")  # only misclassification matrix
plot_vacalib(vacalib_fit = vacalib_eava, toplot = "csmf")  # only CSMF estimates
plot_vacalib(vacalib_fit = vacalib_eava, toplot = "both")  # both (default)
\end{verbatim}

This similarly applies to InSilicoVA, InterVA, and ensemble VA-Calibration. For example, Figure \ref{fig:comsamoz-ensemble-plot} shows the plot returned for the ensemble calibration.

\subsection{Multi-country calibration: Application to CA CODE}\label{implement-cacode}

The Child and Adolescent Causes of Death Estimation (\href{https://childmortality.org/about}{CA CODE}) initiative exemplifies the growing global effort to generate reliable, comparable, and policy-relevant CSMF estimates across countries \citep{unigme2026, wahi2026}. As countries increasingly rely on such estimates to inform national health priorities and resource allocation, robust methodological frameworks are needed to harmonize data sources of varying quality and completeness. VA data, collected widely in LMICs, provide critical inputs for these efforts but are subject to algorithm- and country-specific misclassification biases. Multi-country VA-Calibration addresses this by systematically adjusting for structured patterns and cross-country heterogeneity in the CCVA misclassification. The resulting uncertainty-quantified calibrated CSMF estimates are both locally relevant and globally comparable. Building on the methodological foundation of \href{https://childmortality.org/about}{CA CODE} and its predecessors (\href{https://publichealth.jhu.edu/institute-for-international-programs/our-work/past-projects/child-health-epidemiology-reference-group-cherg}{CHERG} and \href{https://publichealth.jhu.edu/institute-for-international-programs/our-work/past-projects/maternal-and-child-epidemiology-estimation}{MCEE}), this facilitates consistent, scalable, and transparent integration of VA-based mortality evidence across diverse settings to support regional, national, and global estimation frameworks.

The VA data compiled in \href{https://childmortality.org/about}{CA CODE} span multiple countries and provide broad cause-specific death counts across countries, along with relevant covariates. These data are analyzed within a multinomial regression framework, enabling the assessment of covariate effects on CSMFs for countries lacking direct VA data. Since the observed VA-based death counts are subject to CCVA misclassification, the \href{https://cran.r-project.org/package=vacalibration}{\texttt{vacalibration}} package serves as a critical complement to this framework, producing calibrated CSMF estimates and death counts for use in subsequent modeling \citep{mulick2022}.

The top panel in Table \ref{tab:cacode-data} provides an illustrative example where neonatal death counts from five studies in \href{https://childmortality.org/about}{CA CODE} are observed across various countries. This introduces three key distinctions compared to the illustration described in Section \ref{odk-to-ccva}-\ref{implement-comsamoz} from \href{https://comsamozambique.org/}{COMSA-Mozambique}:

\begin{itemize}

\item CCVA algorithms (e.g.~PCVA) may differ from EAVA, InSilicoVA, and InterVA, for which CCVA misclassification matrix estimates are available.

\item Observed broad causes (``congenital'', ``diarrhoea'', ``intrapartum'', ``other'', ``pneumonia'', ``preterm'', ``sepsis'', ``tetanus'') may not align with the broad causes in CHAMPS for which CCVA misclassification matrix estimates are available.

\item Certain causes may be unobserved (or missing), and they may differ across studies.

\end{itemize}

\paragraph{Matching CCVA algorithms.}

To apply CHAMPS-based CCVA misclassification matrix estimates for calibrating other CCVA algorithms, we map each algorithm listed in Table \ref{tab:cacode-data} to the most similar mechanistic approximation found in \texttt{CCVA\_missmat}. Death counts based on death certificate diagnoses are considered the gold standard and, therefore, are not calibrated. Misclassification estimates are specifically available for InSilicoVA and InterVA. While estimating a dedicated misclassification matrix for PCVA would be ideal, we note that the underlying methodology of physician-coded verbal autopsy (PCVA) closely resembles that of EAVA. Consequently, we use EAVA's misclassification matrix to calibrate death counts derived from PCVA.

\paragraph{Matching broad causes.}

As with algorithm alignment, the use of misclassification estimates also requires mapping study-specific causes to the CHAMPS broad causes. In \texttt{vacalibration()}, this is done via the \texttt{studycause\_map} argument. For this example, we define the following mapping based on medical guidance:

\begin{verbatim}
studycause_map_cacode = c("congenital"="congenital_malformation",
                          "diarrhoea"="sepsis_meningitis_inf",
                          "intrapartum"="ipre", "other"="other",
                          "pneumonia"="pneumonia", "preterm"="prematurity",
                          "sepsis"="sepsis_meningitis_inf", 
                          "tetanus"="sepsis_meningitis_inf")
\end{verbatim}

\noindent When some causes in a study are unobserved, we subset the misclassification matrix for the observed causes, and then normalize row-wise.

\paragraph{VA-Calibration of death counts.}

For calibrating death counts in the top panel of Table \ref{tab:cacode-data}, they are input via \texttt{va\_data} and with study cause matching input as \texttt{studycause\_map=studycause\_map\_cacode}. Below, we present the implementation.

\begin{verbatim}
######## load library
library(vacalibration)

######## cacode data
df_cacode = 
  rbind.data.frame(c("India", "PCVA", 48, 139, 29, 9, 38, 10, 17, 9),
                   c("Nepal", "PCVA", 67, 30, 3, 43, NA, NA, NA, 3),
                    c("Congo", "Death Certificate", 20, 28, 1, 5, NA, NA, 2, 0),
                   c("Indonesia", "InSilicoVA", 36, 58, NA, 7, 5, NA, NA, 3),
                   c("Ethiopia", "InterVA", 31, 14, 5, 9, 34, 2, NA, 95))
causes_cacode = c("intrapartum", "preterm", "congenital",
                  "sepsis", "pneumonia", "diarrhoea", "tetanus", "other")
colnames(df_cacode) = c("Country", "Algorithm", causes_cacode)

######## algorithm matching
vaalgo_map_cacode = c("PCVA" = "eava", "PCVA" = "eava", "Death Certificate" = NA,
                      "InSilicoVA" = "insilicova", "InterVA" = "interva")

######## cause matching
studycause_map_cacode = c("congenital"="congenital_malformation",
                          "diarrhoea"="sepsis_meningitis_inf",
                          "intrapartum"="ipre", "other"="other",
                          "pneumonia"="pneumonia", "preterm"="prematurity",
                          "sepsis"="sepsis_meningitis_inf", 
                          "tetanus"="sepsis_meningitis_inf")

######## calibrating
df_cacode_calibrated = df_cacode
vacalib_fit = vector(mode = "list", length = nrow(df_cacode))
for(i in 1:nrow(df_cacode)){
  
  # the study is not calibrated if algorithm is matched with NA
  calibrate = !is.na(vaalgo_map_cacode[df_cacode$Algorithm[i]])
  
  if(calibrate){
    
    # va_data for for a study
    va_data_i = as.numeric(df_cacode[i,causes_cacode])
    names(va_data_i) = causes_cacode  # cause-specific death counts
    
    # preparing va_data with observed causes
    nonNA_indx = !is.na(va_data_i)  # cause id's with observed death counts
    va_data_i = list(va_data_i[nonNA_indx])
    names(va_data_i) = vaalgo_map_cacode[df_cacode$Algorithm[i]]
    
    # fitting VA-Calibration
    vacalib_fit[[i]] = vacalibration(va_data = va_data_i,
                                     age_group = "neonate", 
                                     country = df_cacode$Country[i],
                                     studycause_map = studycause_map_cacode)
    
    # calibrated death counts
    df_cacode_calibrated[i,causes_cacode[nonNA_indx]] = 
      vacalib_fit[[i]]$va_deaths_calib_algo[1,]
    
  }
  
}
\end{verbatim}

\begin{figure}[!t]

{\centering \includegraphics[width=0.95\linewidth]{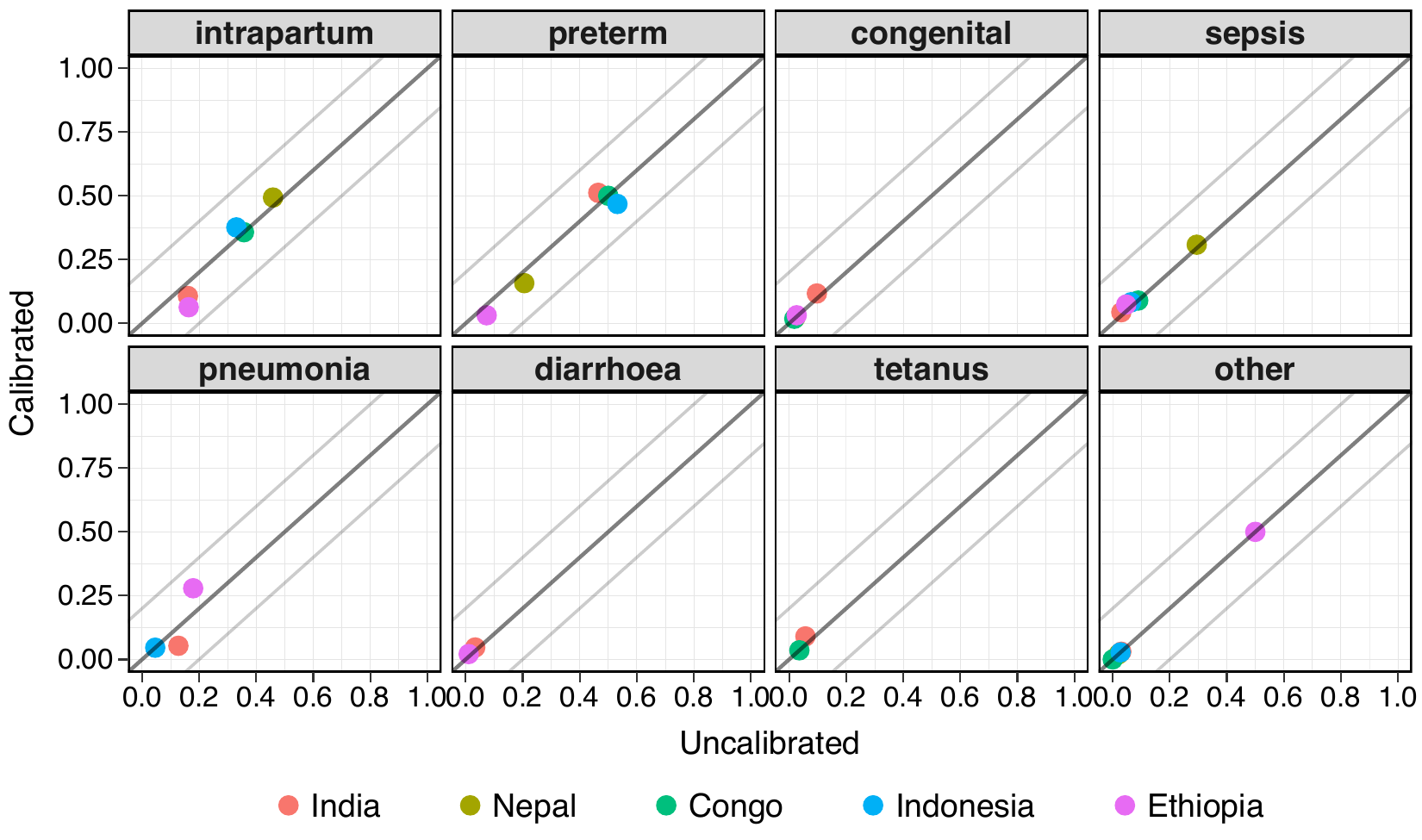} 

}

\caption{Comparison of uncalibrated and calibrated estimates of cause-specific mortality fractions (CSMFs) across five studies in the CA CODE example, as presented in Table \ref{tab:cacode-data}.}\label{fig:cacode-plot}
\end{figure}

\noindent The bottom panel of Table \ref{tab:cacode-data} presents the calibrated death counts, \texttt{df\_cacode\_calibrated}. Figure \ref{fig:cacode-plot} summarizes the table by comparing the uncalibrated and calibrated CSMF estimates derived from the counts.

\begin{table}
\centering
\caption{\label{tab:cacode-data}Neonatal deaths by eight broad causes in five representative CA CODE studies}
\centering
\resizebox{\ifdim\width>\linewidth\linewidth\else\width\fi}{!}{
\begin{tabular}[t]{l|l|l|l|l|l|l|l|l|l}
\noalign{\hrule height 1pt}
\textbf{Country} & \textbf{Algorithm} & \textbf{intrapartum} & \textbf{preterm} & \textbf{congenital} & \textbf{sepsis} & \textbf{pneumonia} & \textbf{diarrhoea} & \textbf{tetanus} & \textbf{other}\\
\noalign{\hrule height 1pt}
\hline
\multicolumn{10}{l}{\textbf{Uncalibrated:}}\\
\hline
\hspace{1em}India & PCVA & 48 & 139 & 29 & 9 & 38 & 10 & 17 & 9\\
\hline
\hspace{1em}Nepal & PCVA & 67 & 30 & 3 & 43 &  &  &  & 3\\
\hline
\hspace{1em}Congo & Death Certificate & 20 & 28 & 1 & 5 &  &  & 2 & \vphantom{1} 0\\
\hline
\hspace{1em}Indonesia & InSilicoVA & 36 & 58 &  & 7 & 5 &  &  & 3\\
\hline
\hspace{1em}Ethiopia & InterVA & 31 & 14 & 5 & 9 & 34 & 2 &  & 95\\
\noalign{\hrule height 1pt}\\
\hline
\multicolumn{10}{l}{\textbf{Calibrated:}}\\
\hline
\hspace{1em}India & PCVA & 32 & 153 & 35 & 13 & 16 & 14 & 27 & 9\\
\hline
\hspace{1em}Nepal & PCVA & 72 & 23 & 3 & 45 &  &  &  & 3\\
\hline
\hspace{1em}Congo & Death Certificate & 20 & 28 & 1 & 5 &  &  & 2 & 0\\
\hline
\hspace{1em}Indonesia & InSilicoVA & 41 & 51 &  & 9 & 5 &  &  & 3\\
\hline
\hspace{1em}Ethiopia & InterVA & 12 & 6 & 6 & 14 & 53 & 4 &  & 95\\
\noalign{\hrule height 1pt}
\end{tabular}}
\end{table}

\section{Discussion}\label{discussion}

In this article, we introduced the \href{https://cran.r-project.org/package=vacalibration}{\texttt{vacalibration}} R package. Using a modular Bayesian framework, it corrects for the misclassification of CCVA algorithms when estimating population-level CSMF estimates from VA data. The primary goal of this work is to provide practitioners with an accessible and reproducible workflow for calibrating VA-only data, thereby enhancing the accuracy of mortality estimates used for policy and surveillance.

The package is implemented in \texttt{rstan}, and enables rapid computation, typically within seconds to minutes depending on the number of broad causes and algorithms. The framework is a fully Bayesian hierarchical model and thus provides uncertainty quantification by default. The modular architecture allows calibration without requiring direct access to the labeled data (such as from \href{https://champshealth.org/}{CHAMPS}), protecting data privacy.

This article focuses on functionalities of the \href{https://cran.r-project.org/package=vacalibration}{\texttt{vacalibration}} package. It relies on CCVA misclassification matrix estimates produced by \citet{Pramanik2026bmjgh} from labeled COD data collected in \href{https://champshealth.org/}{CHAMPS}. Available for three major CCVA algorithms (EAVA, InSilicoVA, and InterVA), two age groups (neonates aged 0-27 days and children aged 1-59 months), and across countries (specific estimates for Bangladesh, Ethiopia, Kenya, Mali, Mozambique, Sierra Leone, and South Africa, and a combined estimate for all other countries), and the estimates are publicly available on the \href{https://github.com/sandy-pramanik/CCVA-Misclassification-Matrices}{CCVA-Misclassification-Matrices} GitHub repository and are also included in this package. For a given country, age group, and algorithm, the package automatically retrieves the corresponding misclassification estimates and obtains calibrated CSMF estimates. It also performs ensemble calibration when multiple CCVA outputs are supplied, extending its flexibility.

Two real-world applications illustrate the package's practical utility. The first application in \href{https://comsamozambique.org/}{COMSA-Mozambique} is an instance of single-country calibration, obtaining calibrated CSMFs from raw VA survey records. This streamlined pipeline highlights the package's compatibility and its seamless integration with \href{https://cran.r-project.org/package=openVA}{\texttt{openVA}} ecosystem and existing workflows. The second application in \href{https://childmortality.org/about}{CA CODE} is an instance of multi-country calibration, calibrating summary-level data (broad-cause-specific death counts) and combining VA data from different countries and algorithms. Here, VA-Calibration serves as a vital correction step within a regression framework for global mortality estimation. This ensures that algorithmic misclassification patterns are appropriately accounted for before cross-country modeling. Overall, \href{https://cran.r-project.org/package=vacalibration}{\texttt{vacalibration}} package offers a transparent, scalable, and statistically principled approach for adjusting misclassification of CCVA algorithms in VA-based population-level CSMF estimation. By leveraging CHAMPS-based misclassification information and embedding it within a modular Bayesian framework, the package bridges the gap between labeled and unlabeled mortality data.

Continuous development is underway for this package. The \href{https://github.com/sandy-pramanik/CCVA-Misclassification-Matrices}{CCVA misclassification matrix estimates} will be updated and possibly extended to more age groups and countries as new labeled data from \href{https://champshealth.org/}{CHAMPS} become available. To enhance the accessibility and practical utility of the package, we are developing an interactive web application that wraps the core \texttt{R} functionality in a user-friendly Graphical User Interface (GUI). This development aims to provide practitioners with easy analysis options without requiring deep proficiency in \texttt{R} or statistical programming, and ensuring that accurate and reproducible CSMF estimates can be generated and applied by a wider global health community. Future developments will incorporate additional methods for uncertainty propagation, particularly those that cut off feedback from unlabeled to labeled data \citep{song2025nevicut}. The framework is also extensible to accommodate next-generation CCVA algorithms, including machine learning and large language model-based approaches, which are expected to improve individual-level accuracy \citep{chen2025lava, chu2025llmmlva}. Future work will focus on extending VA-Calibration to account for cross-country heterogeneity in multi-cause calibration \citep{fiksel2022} and time-varying misclassification. As global mortality surveillance increasingly relies on VA systems, tools like the \href{https://cran.r-project.org/package=vacalibration}{\texttt{vacalibration}} package will be essential to ensure that population-level mortality estimates are accurate, reproducible, and actionable for policy and research.

Beyond its application to COD estimation that we discuss here, the methodological principles underlying VA-Calibration have broad relevance across scientific domains where data-generating mechanisms are imperfectly known or inherently noisy. Misclassification is a pervasive challenge in modern data analysis, extending well beyond global health contexts. For example, statistical inference in association studies using electronic health records (EHR) often experiences outcome misclassification due to incomplete or inaccurate documentation \citep{sinnott2014, Beesley2022}. In the rapidly evolving fields of machine learning and artificial intelligence, algorithmic predictions, whether from rule-based systems or large language models, are seldom perfectly aligned with ground truth. In this broader context, the package can be applied to calibrate population-level prevalence from a discrete classifier or its ensemble using user-supplied fixed or uncertainty-quantified misclassification matrices. This offers a transparent and modular approach that can accommodate a wide range of use cases beyond VA analysis.

\section{Funding}\label{funding}

S.P., E.W., H.K., and A.D. acknowledge funding from the Bill and Melinda Gates Foundation grants INV-034842 and INV-070577. A.A. was also supported by INV-070577. The 2024 Johns Hopkins Data Science and AI Institute Demonstration Projects Award supported S.P., E.W., A.A., and A.D. Additionally, S.P. received support from the Eunice Kennedy Shriver National Institute of Child Health K99 NIH Pathway to Independence Award 1K99HD114884-01A1. H.K., R.B., L.L., J.P., and A.D. were partially supported by funding from the Gates Foundation grant INV-038624.

\bibliographystyle{apalike}
\bibliography{RJreferences.bib}

\end{document}